\documentclass{ws-ijmpe}
\usepackage[super,compress]{cite}
\begin{document}

\markboth{D. Lacroix et al}{Microscopic description of large amplitude collective motion in nuclei}

\catchline{}{}{}{}{}

\title{MICROSCOPIC DESCRIPTION OF LARGE AMPLITUDE COLLECTIVE MOTION IN THE NUCLEAR ASTROPHYSICS
CONTEXT}

\author{DENIS LACROIX and YUSUKE TANIMURA}

\address{Institut de Physique Nucl\'eaire, IN2P3-CNRS, Universit\'e Paris-Sud, \\
F-91406 Orsay Cedex, France \\
lacroix@ipno.in2p3.fr\\
tanimura@ipno.in2p3.fr}

\author{GUILLAUME SCAMPS}

\address{Department of Physics, Tohoku University, \\
Sendai 980-8578, Japan\\
scamps@nucl.phys.tohoku.ac.jp}

\author{C\'EDRIC SIMENEL}

\address{Department of Nuclear Physics, 
Research School of Physics and Engineering \\ 
Australian National University, Canberra, Australian Capital Territory 2601, Australia \\
cedric.simenel@anu.edu.au}
\maketitle

\begin{history}
\end{history}

\begin{abstract}
In the last 10 years, we have observed an important increase of interest in the application 
of time-dependent energy density functional theory (TD-EDF). This approach allows to treat nuclear structure and nuclear reaction from 
small to large amplitude 
dynamics in a unified framework. The possibility to perform unrestricted three-dimensional simulations using state of the art effective interactions
 has opened new perspectives. In the present article, an overview of applications 
where the predictive power of TD-EDF has been benchmarked is given. A special emphasize is made on processes  that are of  astrophysical interest. Illustrations discussed here include giant resonances, fission, binary and ternary collisions leading to 
fusion, transfer and deep inelastic processes. 
\end{abstract}

\keywords{Nuclear density functional theory; transport theory, collective motion.}

\ccode{PACS numbers: 21.60.Jz, 24.10.-i, 24.10.Cn, 25.70.J}


\section{Introduction}
The composition of the universe as we know it today is the result of 
nucleosynthesis. It involves several processes including phenomena where atomic nuclei 
encounter large amplitude collective motion (LACM). Some 
processes involve one or more nuclei to form new isotopes with larger masses and charges. 
Other processes depopulate regions of the nuclear chart. 
In particular, heavy nuclei may encounter fission into
lighter systems. 
\begin{figure}[th]
\begin{center}
\includegraphics[width=12.cm]{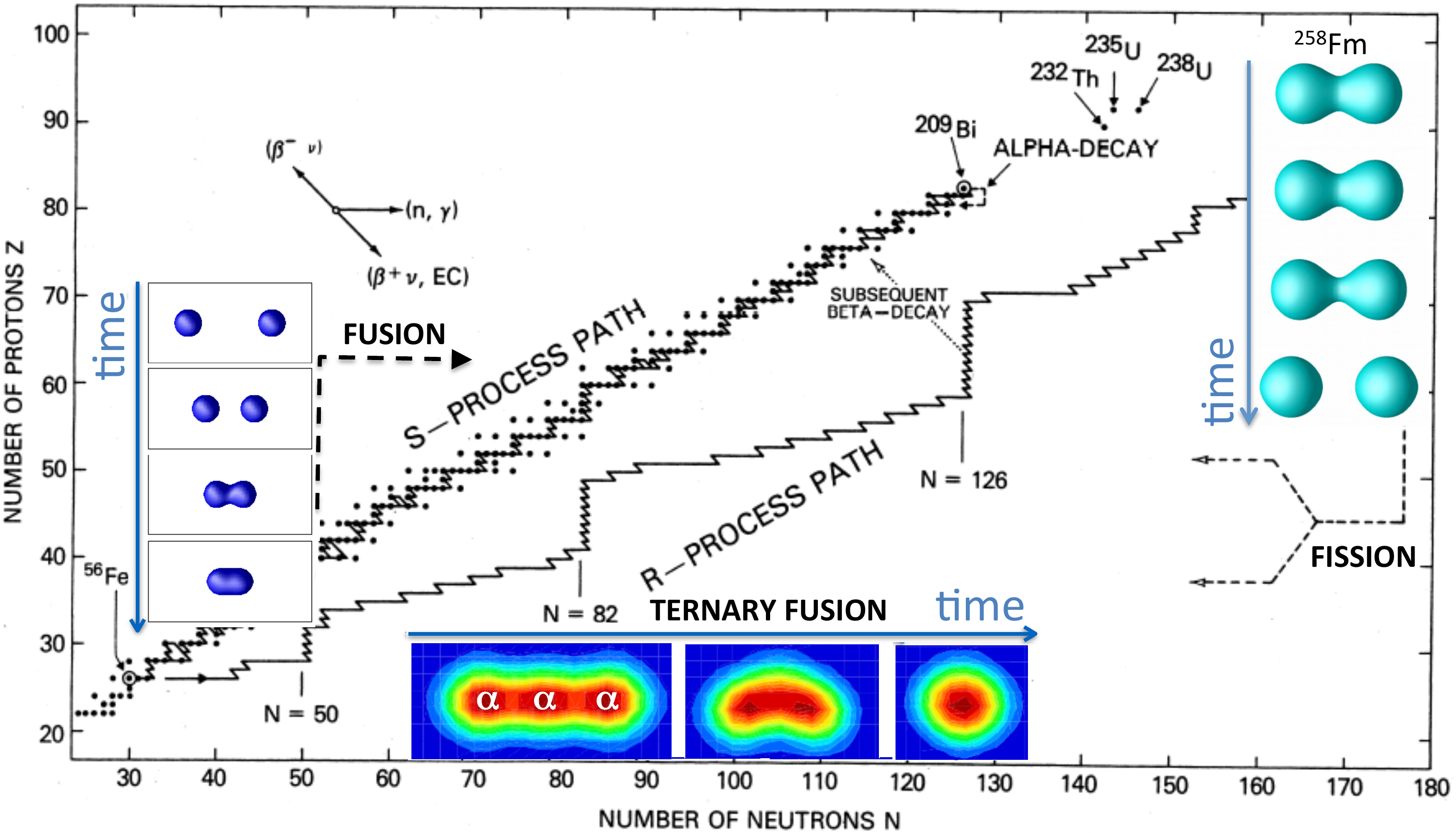}
\caption{\label{fig:gen} (Color online) Schematic view of the nucleosynthesis process (adapted from \cite{Rol88}). 
Examples of large amplitude processes involved in the nucleosynthesis include: the binary fusion process (left), the ternary fusion process (bottom)
and the fission process (right). In the three cases, the density profile at different time have been obtained with the Time-Dependent Energy 
Density Functional approach and are respectively taken from \cite{Sim13a} (binary fusion of two $^{16}$O nuclei), \cite{Uma10} (ternary fusion of three $\alpha$'s) and 
\cite{Sca15} (fission of $^{258}$Fm). 
}
\end{center}
\end{figure}
Several phenomena involving LACM have thus been clearly identified has playing a key role in nucleosynthesis. 
Some of these processes 
are illustrated in Fig. \ref{fig:gen}. 
Amongst the basic nucleosynthesis processes, the pp chain is responsible for most of our Sun energy. This pp chain can however only 
explain the formation of nuclei up to the unstable $^{8}$Be nucleus. As first guessed by F. Hoyle, a bridge between the pp
chain and the CNO cycle can only be understood by invoking the fusion of  three alpha's (ternary fusion) with the intermediate 
formation of a resonance in $^{8}$Be (shown in the lower left side of Fig.  \ref{fig:gen}). 
Similarly, the symmetric complete of partial fusion of light nuclei plays an important role in the formation of nuclei with mass $A < 60$ 
(see left side of Fig. \ref{fig:gen}).  Another example of LACM is the fission process that prevents the formation of heavy- and/or super-heavy elements through the s-process and/or rp-process.

Some of the LACM of astrophysical interest can be directly studied like selected symmetric fusion of light stable 
nuclei. Other cases, like the ternary fusion or reactions involving too exotic nuclei cannot be directly experimented on Earth.  
Microscopic theories able to describe precisely large amplitude phenomena can then give important insights into these reactions.
At present, the only theory that is able to give a unified satisfactory 
framework of all nuclei except the lightest ones, as well as nuclear thermodynamic 
 is the energy density functional (EDF). In the present article, we present an overview of recent progresses 
and successes made in the description of nuclear LACM using the Time-Dependent Energy Density Functional (TD-EDF). 

\section{The Time-Dependent Energy Density Functional theory}

The Time-Dependent Energy Density Functional theory is a generalization of the static EDF \cite{Ben03,Rin80}.  
In this approach, the complex many-body dynamical problem of strongly interacting fermions is replaced by 
the simpler problem of independent particle and/or quasi-particles interacting through a common self-consistent mean-field. 
Including pairing, the TD-EDF equation of motion can be written in a similar form as the Time-Dependent Hartree-Fock Bogolyubov (TDHFB)
theory:
\begin{eqnarray}
i \hbar \frac{d}{dt} {\cal R}(t) & = & \left[ {\cal H}({\cal R}) , {\cal R}(t) \right].  \label{eq:tdhfb}
\end{eqnarray}
Here, ${\cal R}(t)$ and ${\cal H}({\cal R})$ denote the generalized density and the self-consistent mean-field, respectively. They are given 
by 
\begin{eqnarray}
{\cal R} (t) = \left( 
\begin{array} {cc}
\rho(t) & \kappa(t) \\
- \kappa^*(t) & 1-\rho^*(t)  
\end{array} 
\right), ~~~~~{\cal H}  = \left( 
\begin{array} {cc}
h(t) & \Delta(t) \\
- \Delta^*(t) & - h^*(t)
\end{array} 
\right).
\end{eqnarray}
The sub-matrices $\rho(t)$ and $\kappa(t)$ are the normal and anomalous 
densities while $h(t)$ and $\Delta(t)$ are the mean-field and pairing fields. 
In TD-EDF, the fields are  defined as functional derivatives of the energy:
$h_{\mu\nu} = \delta \mathcal{E}[\rho,\kappa,\kappa^*] / \delta \rho_{\nu\mu}$, $\Delta_{\mu\nu} = \delta \mathcal{E}[\rho,\kappa,\kappa^*] / \delta \kappa^*_{\mu\nu}$, where the energy $ \mathcal{E}[\rho,\kappa,\kappa^*]$ is a functional of the densities. 
The parameters entering into the energy functional are directly adjusted 
on infinite nuclear matter and finite nuclei properties and therefore the functional approach already takes into account complex internal 
correlations. In that sense, it goes beyond a HF or HFB theory starting from a bare Hamiltonian. While some differences exist, 
the strategy and goals of the nuclear EDF are similar to the ab-initio DFT approach in condensed matter. The connection becomes 
even more evident noting that most practical (TD-)EDF are constructed from zero-range interaction giving an energy functional of the local normal and anomalous densities and eventually their derivatives. In practice, Eq. (\ref{eq:tdhfb}) is solved by introducing a fictitious trial vacuum associated to a complete set of quasi-particle states $| W_\alpha \rangle = (U_\alpha ,V_\alpha)^T$ that evolves through $i\hbar \partial_t | W_\alpha \rangle = {\cal H}({\cal R}) | W_\alpha \rangle$. Altogether, we end up with the nuclear time-dependent EDF sequence:
\begin{eqnarray}
\{ | W_\alpha (t) \rangle \} &\longrightarrow& {\cal R}(t) = \sum_\alpha | W_\alpha (t)  \rangle \langle W_\alpha (t) | 
\longrightarrow {\cal H}({\cal R})\longrightarrow\{ | W_\alpha (t+\Delta t) \rangle \} \longrightarrow \cdots \nonumber
\end{eqnarray}     
When the nucleus is initially in a normal phase, due to the absence of spontaneous symmetry breaking and more specifically of the $U(1)$ 
symmetry associated to particle number, 
the system cannot dynamically become superfluid. The mean-field dynamics is then disconnected from the anomalous sector and the evolution reduces to a set of single-particle state evolutions similar to TDHF. 
 
Applications made at the early stage of TD-EDF, i.e. in the end of the 70's and beginning of the 80's, 
although very promising \cite{Neg82} were quite restricted due to the numerical constraints at that time. 
During more than 25 years, while static EDF has been constantly improved and confronted to experimental observations,
 time-dependent EDF approaches have been almost completely forsaken except in its small amplitude limit around
equilibrium. Recently, several groups have taken the challenge to provide new TD-EDF codes  without spacial symmetries 
 and that use most advanced functionals consistently with nuclear structure 
\cite{Kim97,Sim01,Nak05,Mar05,Uma05, Was08}. Pairing correlations have been 
included by several groups  \cite{Ave08,Eba10,Ste11,Sca12,Has12,Sca13}. With this renewal of interest, an effort 
is made to get a clear view of the TD-EDF predictive power. Theoretical and technical aspects related to mean-field theory with or without pairing correlations can be found in several recent reviews \cite{Lac04,Sim10,Sim12a,Bul13,Lac14}
In the present review, we present through recent illustrations different processes TD-EDF can successfully describe.

\begin{figure}[htbp]
\begin{center}
\includegraphics[width=10.cm]{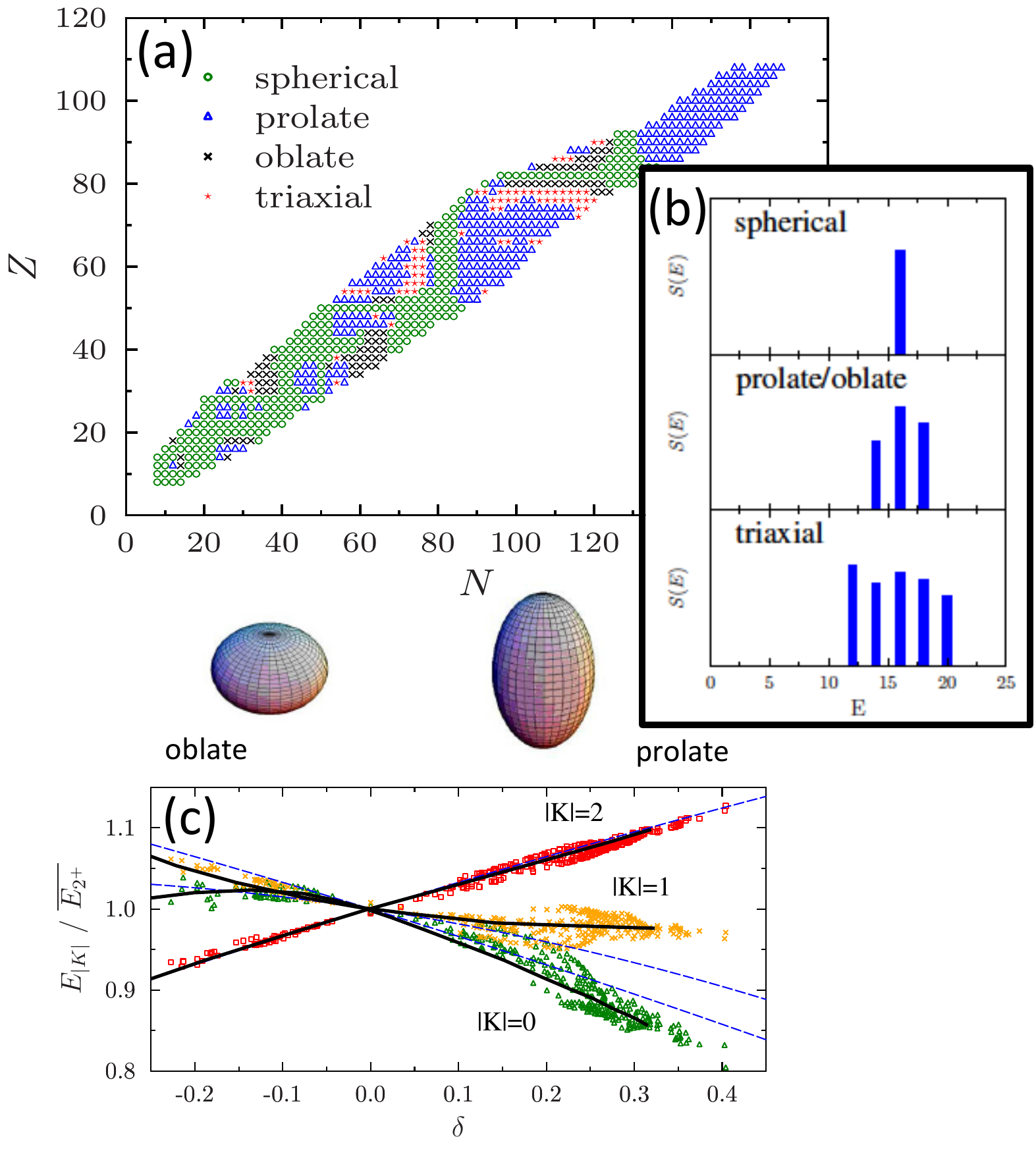}
\caption{\label{fig:coll} (Color online) Illustration of the systematic application of the TD-EDF theory to nuclei including pairing (adapted from \cite{Sca13a,Sca14}). (a) all nuclei that have been considered in the systematic study. Spherical, quadrupole deformed and triaxial nuclei are indicated by different colors. (b) Schematic view of the effect of deformation on the Giant Quadrupole Resonance (GQR). The deformation will induce a fragmentation of the nuclear response. (c) Systematic study of the splitting of collective energy (in quadrupole deformed nuclei) as a function of a deformation parameter $\delta$. }
\end{center}
\end{figure}

\section{From small to large amplitude dynamics in one nucleus}

\subsection{Collective motion in nuclei}
A small external perturbation applied to a nucleus can lead to a cooperative motion of individual nucleons leading to low-lying collective vibrations as well as giant resonances.
The standard way to describe microscopically such ordered motion is to linearize the TD-EDF equation of motion around the 
ground state, leading to the so-called Random-Phase Approximation (RPA) or Quasi-particle RPA (QRPA) \cite{Rin80, Har01,Paa07} .  

This approach is nowadays well under control \cite{Yos06,Hag04,Per08,Pen09,Per11,Paa03,Ter04,Los10,Yos13,Nik13,Lia13}. However, the description of nuclei with various shapes, in particular triaxial deformation remains a technical challenge.  
The recent inclusion of pairing together with unrestricted 3D calculations allow to simulate collective motion 
by direct time evolution on a large scale across the nuclear chart. Indeed, pairing correlations are essential to get realistic properties for the ground state of nuclei especially for open-shell 
systems. The newly developed TD-EDF codes especially including pairing in the BCS approach are very attractive in terms of computational time \cite{Eba10,Sca13}. One important conclusion from these calculations is that TD-EDF results coincide with their 
QRPA counterparts.
Time-dependent approaches have recently been used to make systematic studies like the onset of soft dipole 
mode in exotic nuclei \cite{Eba14} or the influence of deformation on the isoscalar GQR  \cite{Sca13a,Sca14}.   An illustration of results obtained from the systematic analysis made in Ref. \cite{Sca14} is given in Fig.  \ref{fig:coll}.
In that case, more than 700 nuclei were considered. Due to deformation, the response of the nucleus to an external stress is expected to 
present a fragmentation that is characteristic of this deformation. In Fig. \ref{fig:coll}-c, the dependence of the splitting of collective energy has been precisely studied and interpreted.    

\subsection{Fission dynamic}

One obvious advantage of the time-dependent simulation is that, contrary to the linear response theory, it can be applied 
to situations with strong external perturbations or when the nucleus encounters very large deformations. 
\begin{figure}[th]
\begin{center}
\includegraphics[width=11.cm]{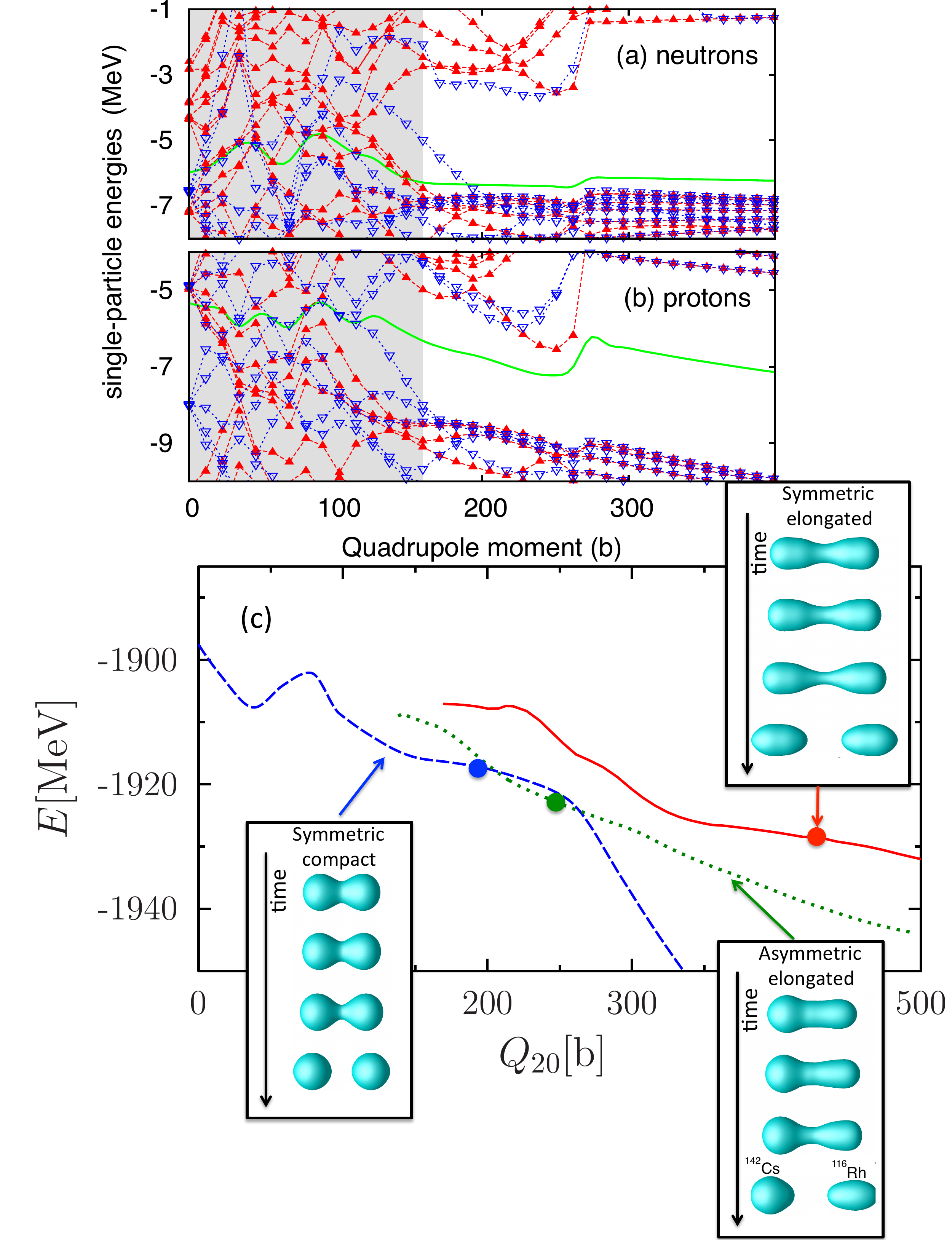}
\caption{\label{fig:fission}(Color online) Simulation of different fission paths in $^{258}$Fm with TD-EDF including pairing. Evolution of neutron (a) and proton (b) single-particle energies along the adiabatic path (symmetric compact case). The red closed and blue open triangles correspond respectively to positive and negative parity states. The thick green lines represent the Fermi energies. 
(c) Adiabatic surfaces corresponding respectively to three different 
paths: symmetric compact (blue dashed line), symmetric elongated (red solid line) and  asymmetric elongated (green dotted line). The three insets correspond respectively to three non-adiabatic  evolutions obtained using TD-EDF. In the three cases, the starting point 
corresponds to the filled circles indicated in the corresponding adiabatic curve. See Ref.\cite{Sca15} for details.}
\end{center}
\end{figure}
For instance, when the strength of the external field is in the non-linear regime one could a priori study anharmonic effects, coupling between different phonons as well as the onset of multiphonons \cite{Cho95,Sim03,Rei07,Sim09}. Larger amplitude motion in heavy systems like the fission process can be studied using the TD-EDF. As pointed out soon after the introduction of this approach in nuclear physics \cite{Neg82,Neg89},  the fission of atomic nuclei is certainly one of the most complex problem to describe in a many-body fermionic interacting system\cite{Neg78}. This stems from the coexistence of quantum effects in both single-particle and collective space (see for instance the recent review\cite{Lac15}).   The description of fission passes through the proper treatment of quantum tunneling in many-body system, spontaneous symmetry breaking, non-adiabatic effects,... In spite of this complexity, this 
problem has recently been revisited in a recent series of works\cite{Sim14,God14, God15, Sca15, Tan15} . An illustration of different fission paths 
in $^{258}$Fm is shown in Fig. \ref{fig:fission} (adapted from Refs \cite{Sca15,Tan15}). A clear advantage of quantal microscopic 
transport theory is that they consistently include nuclear structure and dynamical effects. 
In particular, they do not rely on the assumption that the motion is adiabatic. 

\begin{figure}[htbp]
\begin{center}
\includegraphics[width=12.cm]{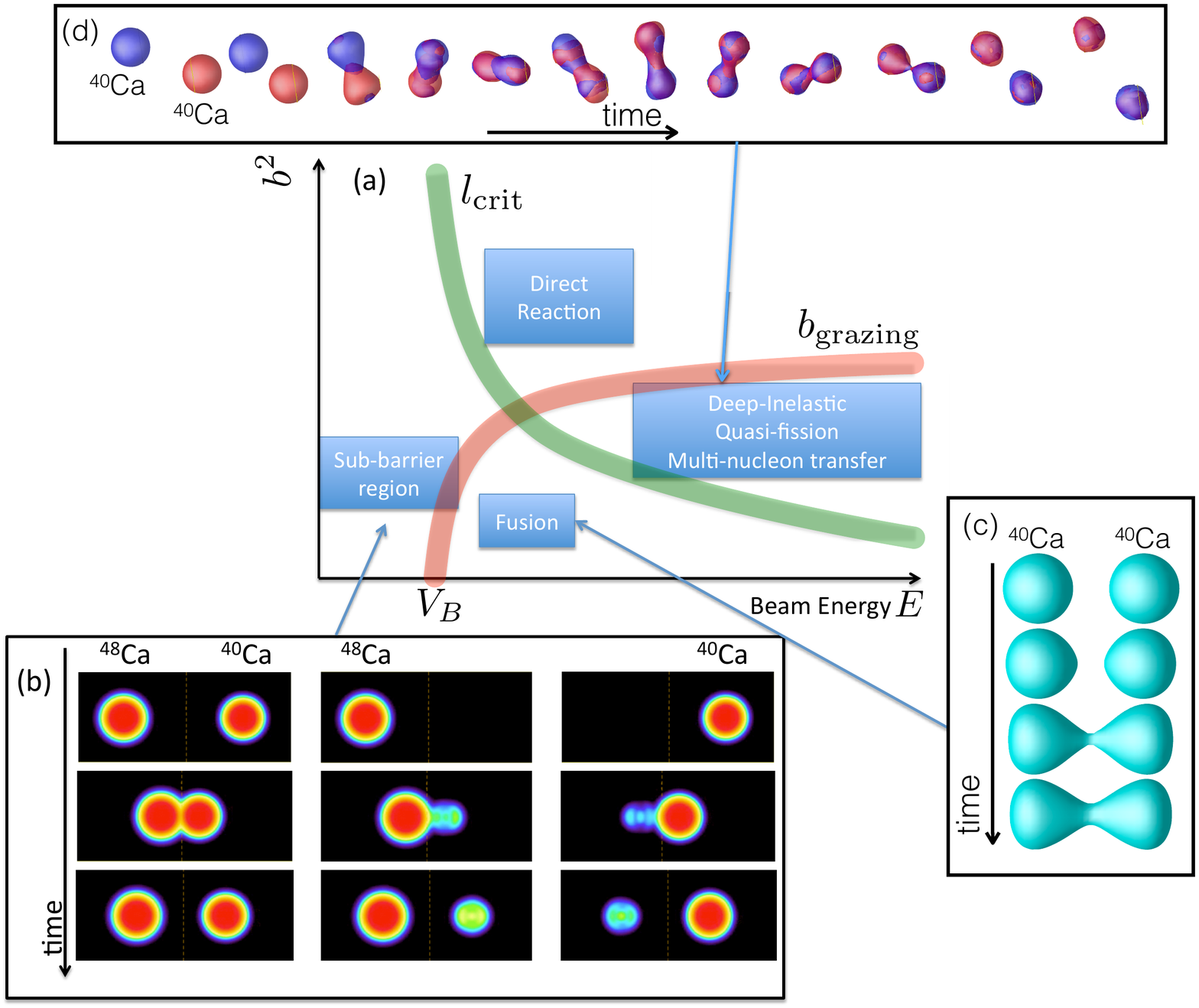}
\caption{\label{fig:binary} (Color online) (a) Simplified view of the different reactions type for energies around the Coulomb barrier as a function of the impact parameter and beam energy. The two lines indicate respectively the grazing impact parameter and the critical angular momentum that can be deposited into a system without fissioning. $l$ and $b$ can be connected through the approximate formula $\hbar(l+1/2)\approx b\sqrt{2\mu E}$ where $E$ is the beam energy and $\mu$ is the reduced mass. (b) Microscopic simulation of transfer reaction below the Coulomb barrier (studied in Ref. \cite{Sca13}). The considered reaction is $^{48}$Ca+$^{40}$Ca at beam energy $E=48.6$ MeV. Left: the density profile of both nucleons are shown before, during and after contact. Middle (resp. Right): the density of nucleons initially in the $^{48}$Ca (resp. $^{40}$Ca) are shown for these three times. (c) Example of fusion in the reaction $^{40}$Ca+$^{40}$Ca (adapted from Ref. \cite{Sim13}). (d) Example of deep-inelastic collision obtained with TD-EDF at non-zero impact 
parameter for the reaction $^{40}$Ca +$^{40}$Ca (see \cite{Sim11}).}
\end{center}
\end{figure}

Still, one difficulty is that mean-field approaches cannot 
describe the dynamics from a very compact shape of the initial nucleus to the separated outgoing nuclei. This stems 
 from (i) the quasi-classical nature of collective dynamics leading to the absence of tunneling (ii) the wrong treatment  of Landau-Zener effects close to single-particle energy crossing. The latter problem is seen both if dynamical pairing is included \cite{Sca15} or not 
 \cite{God14,God15}. To overcome this limitation, it is assumed that the system follows a simple adiabatic dynamics from a compact shape to some already elongated shape outside the barrier. Then, it is possible to perform TD-EDF evolution starting from the latter deformed configuration\cite{Neg78}. This approach has been shown to lead to total kinetic energies and masses of nuclei after fission compatible with experimental observations \cite{Sim14,God15,Sca15}.  These results open new perspectives towards a fully microscopic understanding 
 of the fission process in nuclei.
 

\section{Binary collisions}

In the previous sections, we have shown illustrations of small and large amplitude collective motions where 
a single nucleus is initially considered. Since most nuclear properties can only be uncovered through 
nuclear reactions involving Heavy-Ion accelerators,  a clear condition to access these properties
is a precise understanding of the nuclear reaction itself. Experimentally, only initial and final products 
of the reaction can be known and/or detected.  Due to the inclusion of single-particle quantum effects, 
the TD-EDF approach offers the possibility to describe many facets of low energy nuclear reactions, typically with beam energies 
lower than 10-15 MeV/A.  

One of the richness of nuclear physics is the diversity of phenomena that could be observed depending on the beam energy 
and/or impact parameter of the reaction. While other approaches are usually dedicated to specific reaction channels, TD-EDF 
provides a single framework able to account for many of these channels from most peripheral to most central reactions including 
Coulomb and nuclear effects, collective and single-particle intrinsic excitations. Some of the channels of interest 
are shown in Fig. \ref{fig:binary} (panel (a)) as well as some corresponding simulations obtained with TD-EDF.  Below, we give a summary of the recent discussions in the field for selected processes:
\begin{itemize}
  \item {\bf Fusion reaction above the Coulomb barrier:} 
  
  The description of fusion above  the Coulomb barrier is one of the historical successes of TD-EDF \cite{Bon78,Neg82} . The TD-EDF 
  approach corresponds to a semi-classical approximation for the relative motion of colliding nuclei during  the approaching phase and 
  properly accounts for the possible excitation of internal degrees of freedom during the reaction\cite{Sim13}.  Due to this semi-classical nature, 
   fusion can only occur for center of mass energies above the Coulomb barrier and up to an impact parameter that depends on the beam energy. Then, fusion cross-sections can be calculated using the semi-classical approximation:
  \begin{eqnarray}
\sigma_{\rm Fus} (E_{\rm c.m.}) \simeq \frac{\pi\hbar^2}{2 \mu E_{\rm c.m.}} \left[ l_{\rm max}(E_{\rm c.m.}) +1 \right]^2, \nonumber
\end{eqnarray}
 where $\mu$ is the reduced mass,  $E_{\rm c.m.}$ is the center of mass energy and $l_{\rm max}$ is the maximal angular 
 momentum for which fusion takes place (for more details see \cite{Fro96,Sim12}). The estimate of fusion cross-section turns out 
 to reproduce rather well experimental observations. Besides the cross-sections, such calculations give important information on the minimal energy 
  or on the maximal angular momentum at which fusion can occur. In particular, the former quantity was shown to be influenced by 
  dynamical effects leading to a deviation from the limit where the density of nuclei are frozen in the approaching phase \cite{Was08,Uma14}.   
  These dynamical effects have been observed using different techniques leading to the notion of dynamical fusion barrier. 
  
  Besides the location of the fusion barrier, two methods have been used to extract directly the nucleus-nucleus potential at all relative distances, 
  namely the density constrained TDHF (DC-TDHF)\cite{Uma06,Jia14}  and the dissipative-dynamics TDHF (DD-TDHF)\cite{Lac02,Was08}. In particular, even if the sub-barrier fusion cannot be described with the TD-EDF approach, extracting 
  the nucleus-nucleus potential from it allows to estimate the fusion cross-section at all energies (see for instance 
  \cite{Uma07,Uma08,Obe12,Kes12,Des13,Obe13}). In addition to the fusion in the sub-barrier regime, the specific case of heavy and very heavy elements has been considered in \cite{Guo12,Sim12a,Was15} giving  some physical insight in the so-called "extra-push" energy necessary to form these systems. It is finally worth mentioning the specific study of nuclear fusion between light nuclei for astrophysical interest that has been made in Ref. \cite{Uma12}.    
  
   \item {\bf Nucleon transfer in the sub-barrier or deep-sub-barrier regime:} In recent years, clean experiments have been made to 
   provide information on nucleon transfer below the Coulomb barrier \cite{Cor11,Mor14,Eve11}. The TD-EDF framework provides a quantum 
   description of single-particle transfer. An illustration of the single-particle transfer reaction is given in Fig. \ref{fig:binary} (panel (b)).  
   One difficulty is that reaction partners after the reactions correspond to a strongly entangled 
   many-body wave-function. 
   Guided by the particle number projection technique used in the nuclear structure context, a method has been proposed in Ref. \cite{Sim10b}
   to extract probabilities to transfer one or more particles from one nucleus to the other. 
   Recent applications include systematic studies and comparisons with experimental data\cite{Sek13}. 
An interesting extension has also been proposed to extract the excitation energy in the fragments produced in the various transfer channels\cite{Sek14}.
   The particle number projection technique has been generalized to treat the transfer to/from superfluid nuclei in Ref. \cite{Sca12} .
      
     \item {\bf Deep inelastic, quasi-fission and dissipative aspects:} TD-EDF approach contains dissipative effects associated to the reorganization of single-particle states and to the motion of the nuclear shape. In particular, it includes the dissipation due to the nucleon exchange (window mechanism) and the dissipation induced by the resistance of the single-particle states to the change in the self-consistent mean-field (wall effect) \cite{Lac15}. The TD-EDF dynamics can be mapped into a simple one dimensional evolution for the 
relative distance $R(t)$ and relative momentum as:
\begin{eqnarray}
\dot P = -\frac{\partial V(R)}{\partial R} + \frac{1}{2} \mu[R(t)] \dot R^2(t)- \gamma(R) \dot R(t) \nonumber
\end{eqnarray}      
where $\mu$ and $\gamma$ are respectively the reduced mass and friction coefficients. 
This macroscopic mapping that is the basis of the DD-TDHF approach \cite{Lac02,Was08,Was09} not only has given access to the nucleus-nucleus potential as discussed above but also was one of the first attempt to obtain information on dissipation directly from TD-EDF. 
A second approach, based on the interpretation of mean-field dynamics as an average evolution over fluctuating initial conditions 
has also been proposed and applied in Refs \cite{Ayi09,Was09b,Yil11,Yil14}. The extracted transport coefficients turns out to be in good agreement with those usually used in macroscopic-microscopic approaches \cite{Ada97}. 
In the case of small fluctuations, the latter approach could be reduced\cite{Ayi08} to the time-dependent RPA\cite{Bal84}. 
The TDRPA has been applied to compute fragment mass and charge distributions in giant resonances \cite{Tro85,Bro08} and in deep-inelastic collisions \cite{Bon85,Mar85,Sim11}.

More generally, time-dependent mean-field approaches can provide important insight into reaction mechanisms where dissipation occurs in collective space. 
This framework is in particular optimal to provide the mean evolution of single-particle operators like mean particle number, total angular
momentum and/or total kinetic energy. It is therefore particularly suited for the description of semi-peripheral collisions associated to deeply inelastic collisions like multi-nucleon transfer or quasi-fission\cite{Wak14,Obe14,Ham15}.  An illustration of quasi-fission dynamics is shown 
in Fig. \ref{fig:binary}-(d). 
In addition to a good reproduction of experimental quasi-fission fragment mass-angle distributions\cite{Wak14}, 
TD-EDF calculations can be used to investigate
 energy sharing between kinetic energy of 
outgoing fragments and dissipation during the collision\cite{Obe14}.    
\end{itemize}  

\begin{figure}[htbp]
\begin{center}
\includegraphics[width=8.cm]{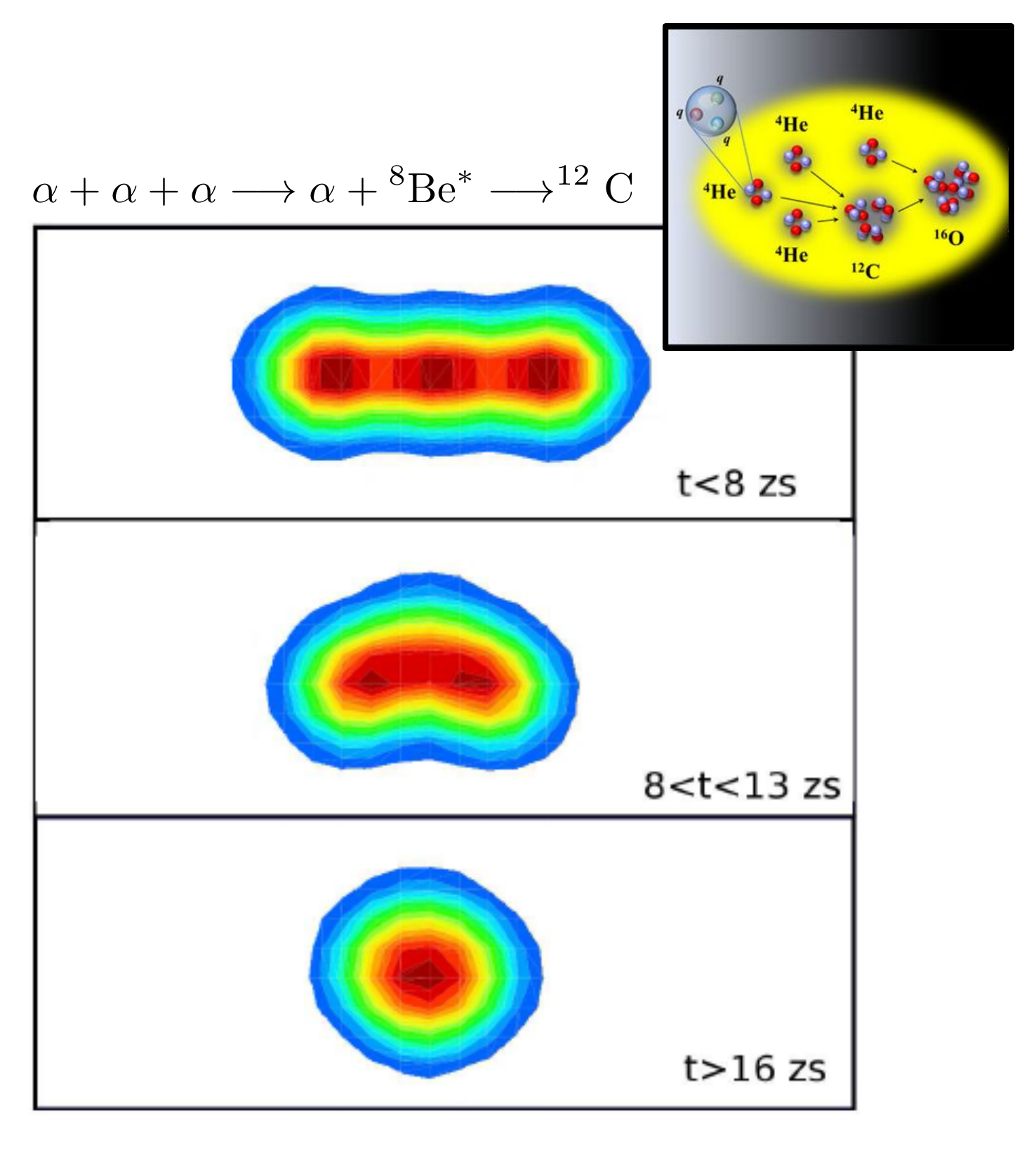}
\caption{(Color online) Ternary fusion described by TD-EDF. Three alpha particles initially in contact are evolved in time to
finally form a $^{12}$C compound nucleus (adapted from \cite{Uma10}).}
\end{center}
\label{fig:ternary}
\end{figure}

\section{Ternary collisions}
As a final example, we want to stress that microscopic mean-field theory can also provide insights into reactions 
that could hardly be tested experimentally on earth. This is the case for reactions involving very exotic nuclei but also with reactions 
involving  more than two nuclei. A typical example of particular astrophysical interest is the 
possible reaction between three $\alpha$ particles leading ultimately to the formation of a $^{12}$C nucleus.  While the importance 
of this process has been proposed by F. Hoyle long ago, very little is known on the way such a reaction can occur dynamically. The process has been recently simulated using TD-EDF in Ref. \cite{Uma10} giving for the first time a possible time-dependent view
of the ternary fusion processes (see Fig. \ref{fig:ternary}). At present, the description of dynamical process using TD-EDF in the nuclear astrophysics context has been only scarcely explored. But, certainly, the study made in Ref. \cite{Uma10} demonstrates that it can bring some interesting 
 insight in this field. 
 \section{Conclusion}
 
 In the present article, we tried to give an overview of the successes of TD-EDF through recent examples of applications. 
 With the renewal of interest in the field, it is more and more evident that the time-dependent microscopic 
 mean-field approach has become an important tool for the description of nuclear small and large amplitude motions. In 
 particular, it turns out to have a rather good predictive power for many physical processes while starting directly from few 
 parameters associated to the effective interactions.  
 TD-EDF approaches are thus well suited to predict nuclear dynamics relevant in the astrophysics context as it often involves very exotic systems for which very little is known. 
For instance, the dynamic of the most exotic systems such as neutron star crusts have been recently investigated with such approaches \cite{Seb09,Seb11,Sch13,Sch14,Sch15}.
 
 Efforts have been made recently to improve the description of quantum fluctuations in collective space 
 and to go beyond the independent particle or quasi-particle approaches. Several promising and tractable approaches have been 
 recently proposed and tested \cite{Sim12a,Lac14,Lac15}. The ultimate goal is to provide a unified description of nuclear structure 
 and reaction processes with high predictive power for low energy nuclear physics.

\section*{Acknowledgements}
This work has been partly supported by the Australian Research Council under Grants No. FT120100760.
G.S. acknowledges the Japan Society for the Promotion of Science
 for the JSPS postdoctoral fellowship for foreign researchers.
 This work was supported by Grant-in-Aid for JSPS Fellows No. 14F04769.

\end{document}